\documentclass[conference]{IEEEtran}
\IEEEoverridecommandlockouts
\usepackage{cite}
\usepackage{amsmath,amssymb,amsfonts}
\usepackage{algorithmic}
\usepackage{graphicx}
\usepackage{textcomp}
\usepackage{xcolor}
\usepackage{hyperref}

\usepackage{comment}

\def\BibTeX{{\rm B\kern-.05em{\sc i\kern-.025em b}\kern-.08em
    T\kern-.1667em\lower.7ex\hbox{E}\kern-.125emX}}

\begin{document}

\title{Breaking Down the Parallel Performance of GROMACS, a High-Performance Molecular Dynamics Software}

\author{\IEEEauthorblockN{Måns I. Andersson, N. Arul Murugan, Artur Podobas, Stefano Markidis}
\IEEEauthorblockA{\textit{Division of Computational Science and Technology},\\ KTH Royal Institute of Technology,\\ Stockholm, Sweden \\
\{mansandes,murugan,podobas,markidis\}@kth.se}
}

\maketitle              
\begin{abstract}
GROMACS is one of the most widely used HPC software packages using the Molecular Dynamics (MD) simulation technique. In this work, we quantify GROMACS parallel performance using different configurations, HPC systems, and FFT libraries (FFTW, Intel MKL FFT, and FFT PACK). We break down the cost of each GROMACS computational phase and identify non-scalable stages, such as MPI communication during the 3D FFT computation when using a large number of processes. We show that the Particle-Mesh Ewald phase and the 3D FFT calculation significantly impact the GROMACS performance. Finally, we discuss performance opportunities with a particular interest in developing GROMACS for the FFT calculations. 
\end{abstract}
\section{Introduction}
Molecular Dynamics (MD)~\cite{karplus2002molecular} is the use of computer simulations to study the physical system particle dynamics and interactions. Today, this technique is widely used in different scientific domains, such as biochemistry and material science, among many others. In particular, the MD software landscape is dominated by a number of well known HPC codes, including GROMACS~\cite{van2005gromacs}, NAMD~\cite{phillips2005scalable}, and CHARMM~\cite{brooks2009charmm}. 

In this work, we investigate the GROMACS parallel performance. GROMACS originated in the early 1990s at the University of Groningen~\cite{berendsen1995gromacs} and has since then been developed and maintained as a community effort. It supports an open-source policy and, among its many strengths, can be executed on a large number of systems, including small (personal) laptops all the way to large high-performance computers (HPC)~\cite{abraham2015gromacs}. Furthermore, GROMACS supports both general-purpose processors (CPUs) as well as Graphics Processing Units (GPUs)~\cite{https://doi.org/10.1002/jcc.26011}. However, despite the continuous improvement in hardware technologies, GROMACS (and other MD frameworks) are still challenged - from the computational point of view - to simulate critical biological processes such as protein folding, conformational transition in bio-molecules (such as R to T transition in Hemoglobin), bacterial and viral infections~\cite{elber2005long,borhani2012future}. These simulations require the usage of supercomputers and accelerators. Needless to say, GROMACS is constantly developed and extended to improve its parallel efficiency as well as algorithmic improvements~\cite{pall2014tackling} to facilitate the study of larger and more complex molecular simulations. 

This paper seeks to understand the GROMACS parallel performance and identify optimization possibilities. For this reason, we run GROMACS on two state-of-the-art HPC systems, analyze their results, and identify key performance-limiting characteristics. In short, our contributions are:
\begin{enumerate}
\item A quantitative and systematic performance evaluation of GROMACS on two HPC systems.
\item We study the impact of different GROMACS phases' implementations, varying the number of processes. We quantify their different impact on the overall GROMACS performance and analyze their respective performance-degrading contribution. For a different number of processes, we identify various optimization opportunities.
\item We quantify the performance impact of using various FFT libraries.
\end{enumerate}
\section{Background}
\label{back1}
MD simulations mimic the dynamics of molecules by solving numerically the equation of motion of particles, using for instance Verlet or leap-frog algorithms. To determine the new position and velocities for each particle, we need to calculate the force acting for each particle. Examples of such forces are the van der Walls and the Coulombic forces. While the cost of particle position and velocity calculations scale with the number of particle under study, a naïve algorithm for the calculation of the forces requires to calculate the contribution to the force for each pair of particle present in the system, making the computation scaling as the square of number of particles present in the system. 

To decrease the computational complexity of force calculations, modern MD algorithms divide the interactions between the molecules into short-range interaction, such the ones from van der Waals interactions, and long-range interactions, such as electrostatic interactions. For instance, van der Waals forces are short-range in nature and decays with distance rapidly. Short-range interactions are only computed for the neighboring particles within a cut-off distance, usually set to 15-20 Å for many applications. The force contribution from particles farther than the cut-off distance is neglected, effectively reducing the interaction of one particle to the closest particles only. However, the electrostatic interactions are long-range in nature and farther particles contribute to the its calculations, still requiring $O(N^2)$ calculations. To solve, this problem, modern MD codes, such as GROMACS, use the PME method. 

\subsection{Particle-Mesh Ewald Computations}
The basic (and here simplified for sake of clarity) strategy of the PME technique is to discretize the simulation domain in uniform computational grid and calculate the charge density for each grid point, for instance using interpolation functions. After the charge density on the grid points is known, the Poisson Equation $\nabla^2 \Phi = - \rho / \epsilon_0$ is solved on the grid for the electrostatic potential $\Phi$ and the electric field (still on the grid points) is calculated from it as $E = - \nabla\Phi$. The electric field information on the grid is transferred to the particle by using again interpolation functions. The PME method use FFT to solve the Poisson equation for electrostatic potential on the grid: we transform first the charge density information to the spectral space, we solve the Poisson equation in the spectral space as algebraic equation (multiplication in spectral space) and then apply an inverse FFT to calculate the potential in the real-space. By working in the spectral spaces, the PME method accounts also for force contributions arising from periodic infinite systems, such as the typical systems studied with MD.  In the PME method, 1D FFT (and the inverse 1D FFT)  requires  $O(N_g\log N_g)$, where $N_g$ is the number of grid points. To calculate the electrostatic potential we use first a 3D FFT on a real data input (the charge density), and after the convolution we apply a 3D FFT to move the potential to the physical space. The way to perform a 3D FFT is by using several 1D FFT and transposing the results. In a parallel setup, the transpose operation requires a large number of messages is the all-to-all communication that scales linearly with the number of processes. Therefore, the all-to-all communication is a potential performance bottleneck~\cite{pall2014tackling}.

\section{Background}
\label{back1}
MD simulations mimic the dynamics of molecules by numerically solving the equation of particles' motion using Verlet or leap-frog algorithms. To determine each particle's new position and velocities, we need to calculate the force acting on each particle. Examples of such forces are the van der Walls and the Coulombic forces. While the cost of particle position and velocity calculations scale with the number of particles under study, a naïve algorithm for the calculation of the forces requires calculating the contribution to the force for each pair of particles present in the system, making the computation scaling as the square of the number of particles present in the system. To decrease the computational complexity of force calculations, modern MD algorithms divide the interactions between the molecules into short-range interactions, such as the ones from van der Waals interactions, and long-range interactions, such as electrostatic interactions. For instance, van der Waals forces are short-range in nature and decay with distance rapidly. Short-range interactions are only computed for the neighboring particles within a cut-off distance, usually 15-20 Å for many applications. The force contribution from particles farther than the cut-off distance is neglected, effectively reducing the interaction of one particle to the closest particles only. However, the electrostatic interactions are long-range in nature, and farther particles contribute to its calculations, still requiring $O(N^2)$ calculations. 

Modern MD codes, such as GROMACS, use the Particle-Mesh Ewald (PME) method to solve this problem. The basic strategy of the PME technique is to discretize the simulation domain in a uniform computational grid and calculate the charge density for each grid point, for instance, using interpolation functions. After the charge density on the grid points is known, the Poisson Equation $\nabla^2 \Phi = - \rho / \epsilon_0$ is solved on the grid for the electrostatic potential $\Phi$ and the electric field (still on the grid points) is calculated from it as $E = - \nabla\Phi$. The electric field information on the grid is transferred to the particle by using again interpolation functions. The PME method use FFT to solve the Poisson equation for electrostatic potential on the grid: we transform first the charge density information to the spectral space, we solve the Poisson equation in the spectral space as an algebraic equation (multiplication in spectral space), and then apply an inverse FFT to calculate the potential in the real-space. The PME method also accounts for force contributions arising from periodic infinite systems, such as the typical systems studied with MD, by working in the spectral spaces.  In the PME method, 1D FFT (and the inverse 1D FFT)  requires  $O(N_g\log N_g)$, where $N_g$ is the number of grid points. We first use a 3D FFT on a real data input (the charge density) to calculate the electrostatic potential. After the convolution, we apply a 3D FFT to move the potential to the physical space.

GROMACS divides the calculations at different parallelization levels, ranging from MPI to OpenMP, CUDA, and CPU vector instructions. At a high level, \textbf{GROMACS uses a pipelined parallelism} with two main phases: the Particle-Particle (PM) and PME calculations. GROMACS allows dividing the MPI processes into PME processes dedicated only to PME and PP phase responsible for all the other calculations, such as computing the particle dynamics and short-range interactions. The two phases can run in parallel and typically on different kinds of resources, such as different nodes, cores, or devices. To finish a GROMACS computational cycle, the PME phase needs to be completed. This synchronization might introduce a delay in the simulation (causing an idle time on the PP processes) and load imbalance if the PP and PME phases are not finishing at the same time. Naturally, the number of PP and PME MPI processes impacts the load balance and the performance. In GROMACS, the choice of the number of processes dedicated to PME and PP calculations can be set by using the command line (via \texttt{-npme} option) or can be set by GROMACS in preparation for a simulation by an auto-tuning tool: \texttt{gmx tune\_pme} (which is not to be confused with the PME tuning done at run-time). GROMACS allocates 1:3 or 1:2 ranks for the PME and PP computations based on the domain if left unspecified by the user. In this study, we set the ratio of PME to PP equal to 1:3.
 
The GROMACS PME performance largely depends on six major components  \cite{shamshirgar2017comparison}. If we focus on PME calculations, we identify the six major phases as:
\begin{enumerate}
  \item \textbf{Redistribution of positions and forces (X/F)} This phase redistributes atoms, parameters, and coordinates before each 3D FFT calculation. 
  \item \textbf{Spread} Using interpolation functions (often called window functions), such as p-th order b-spline, the charges of the particles are distributed on the uniform grid.  
  \item \textbf{1D FFT calculations} The distributed forward and backward 3D FFT is done with a GROMACS specific 1D or 2D FFT factorization. Currently,  GROMACS allows the use of three different FFT libraries when calculating the FFT on CPU, namely FFTW3~\cite{FFTW05}, FFTMKL, FFTPACK~\cite{SWARZTRAUBER198251} for the PME computation. 
  \item \textbf{3D FFT communication} These costs relate to parallel communication performed during the transpose operations during the 3D FFT operations. In GROMACS, this is achieved either by a \texttt{MPI\_Alltoall} or by FFTW transpose operation if the FFTW's 3D library is used on a single node. 
  When the 3D FFT size in the domain x-direction is evenly divisible by the number of PME ranks, a 2D decomposition is used, which requires less communication than a 1D decomposition.
  \item \textbf{Solution of the Electric field} In this step, we perform the calculation of the electrostatic force by differentiating the electrostatic energy. 
  \item \textbf{Gather} The potential (force or energy) is evaluated at the target particles with the same interpolation functions as in the spreading step.
  \item \textbf{Leonard-Jones} Leonard-Jones is a commonly used potential. It is not used in this paper because it is not possible to run this step on GPU systems at the moment.
\end{enumerate}

\section{Related Work}
Given the importance of GROMACS for MD studies, there is a history of benchmarking the throughput of GROMACS. Ref. \cite{GruberChristianC2011Sbol} discusses  optimal GROMACS configuration for a given problem on a given cluster. An additional performance and benchmark analysis on the SuperMUC supercomputer is Ref. \cite{KutznerCarsten2014SotG}. Ref. \cite{pall2014tackling} presents the future of GROMACS development and discusses the limitations of performance due to PME's limited scaling. As the PME and FFT limit the strong scalability, new algorithmic advancements, such as the use of the Fast Multipole Method (FMM) for MD \cite{fmm} are pursued.

\section{Methodology}
\label{perf1}
This work quantifies GROMACS parallel performance using different configurations, HPC systems, and FFT libraries (FFTW, Intel MKL FFT, and FFT PACK). We break down the cost of each GROMACS computational phase and identify non-scalable stages. The performance evaluation uses test cases that are similar to production runs. To better explain the scaling of the different components, we turn load balancing and PME tuning off.  We evaluate the impact of the PME calculations and associated FFTs using two basic configurations, presented in Table \ref{tab:systems1}. The first system, simulating Lysozyme in water, is a relatively small benchmark system in terms of grid points and the number of particles: the 3D grid consists of 44 $\times$ 44 $\times$ 44, and there are 35,000 atoms.  Instead, the second configuration represents a simulation of the Spike protein. In this case, the grid points are 108 $\times$ 144 $\times$ 144, and the number of atoms is 850,000.  The number of particles is 35,000 and 0.85 million approximately for the two configurations. In particular, the viral Spike protein studied here is involved in the interaction with the host cell receptor called hACE-2 and is responsible for the first phase of viral infection and is one of the potential viral targets for developing Covid-19 therapeutics.
\begin{table}
\caption{Specifications for the use cases }
\begin{center}
\begin{tabular}{|l|l|l|l|}
\hline
MD system           & Lysozyme in water  & Spike protein: ACE-2  \\ \hline 
\# atoms            & 35 000    & 0.85 M     \\ \hline
time step [fs]      & 0.002     & 0.002      \\ \hline
domain size [nm]    & 7 $\times $7 $\times $7 &17 $\times $21 $\times $23  \\ \hline
cut-off radii [nm]  & 1         & 1          \\ \hline 
PME grid [nm]      & 0.16      & 0.16      \\ \hline
PME interpolation order & 4          & 4           \\ \hline 
steps (Beskow)            & 5 000   & 5 000      \\ 
\quad \quad (Tetralith) & 100 000 & 100 000 \\ \hline
\end{tabular}
\end{center}
\label{tab:systems1}
\end{table}

To characterize the performance of FFT libraries on a large scale, we evaluate the CPU code on four systems. For CPU evaluations, we use Beskow and Tetralith with CPU FFT libraries (FFTW, MKL, and FFTPACK). 
In addition to two supercomputers, we evaluate the GROMACS performance on GPUs; we use Kebnekaise and NJ with cuFFT. We note that the PP and PME phases are highly intertwined on GPU, and a clear separation of the phases is challenging (for this reason, we limit the study on GPU to the total performance). We summarize the configurations of the systems in Table~\ref{tab:systems}.

\begin{table}
\caption{The hardware architecture of our evaluation platforms.}
\resizebox{\linewidth}{!}{%

\begin{tabular}{|l|l|l|l|l|}
\hline
Name       & CPU                                                                                       & RAM                                                     & GPU         & Compiler Env.    \\ \hline
Beskow     & 2x Xeon E5-2698v3                                                                         & 64 GB                                                   & -           & GCC 10.3, Intel 19.1, HT on, OpenMPI       \\ \hline
Tetralith  & 2x Intel Xeon Gold 6130                                                                   & 96 GB                                                   & -   & GCC 7, HT off,              \\ \hline
Kebnekaise & \begin{tabular}[c]{@{}l@{}}2x Intel Xeon Gold 6132\\ 2x Intel Xeon E5-2690v4\end{tabular} &                 \begin{tabular}[c]{@{}l@{}}192 GB\\ 128 GB\end{tabular} &\begin{tabular}[c]{@{}l@{}} 2 x NVIDIA V100\\ 2, 4 x NVIDIA K80\end{tabular} & GNU 10.3, CUDA 11.3 \\ \hline
NJ       & AMD EPYC 7302P                                                                            &                                                         & 2 x NVIDIA A100 & GNU, CUDA 11.3   \\ \hline
\end{tabular}
}
\label{tab:systems}
\end{table}

All simulations were performed with GROMACS 2021.3. We compile GROMACS using the optimal settings, as advised in the user guide. We build FFTW from the source. In particular, we specify to use single-precision compute. We also enable vectorization by specifying \texttt{GMX\_SIMD=AVX2\_256} on Beskow and \texttt{GMX\_SIMD= AVX\_512} on Tetralith.  We use the GNU compiler collections on all platforms combined with CUDA when GPU is used. MKL's FFT was compiled with the Intel compiler. 

The figures of merit we use in this paper are the total execution time and nanoseconds (ns) per day. This last metric is how many nanoseconds can be simulated within a day of the simulation and represents the total GROMACS throughput. Each simulation is performed ten times on Beskow and five times on Tetralith, showing a high standard deviation. The figures consist of the mean (median) of these simulations. The simulation parameters are: \texttt{-notunepme -dlb no} and for the GPU \texttt{-notunepme -dlb no -nb gpu -update -gpu}. 

Furthermore, to minimize the effects of congestion and impact from the network topology (explained in detail in \cite{10.1109/SC.2018.00030}), every job evaluates all FFT libraries with the same node configuration. This is not done for the Tetralith simulations. On both Beskow and Tetralith, we run with two OpenMP threads per MPI rank, and on Beskow, hyperthreading is turned on by default. 

\section{Results}
\label{res1}
\begin{figure*}[]
    \centering
\includegraphics[width=1\textwidth]{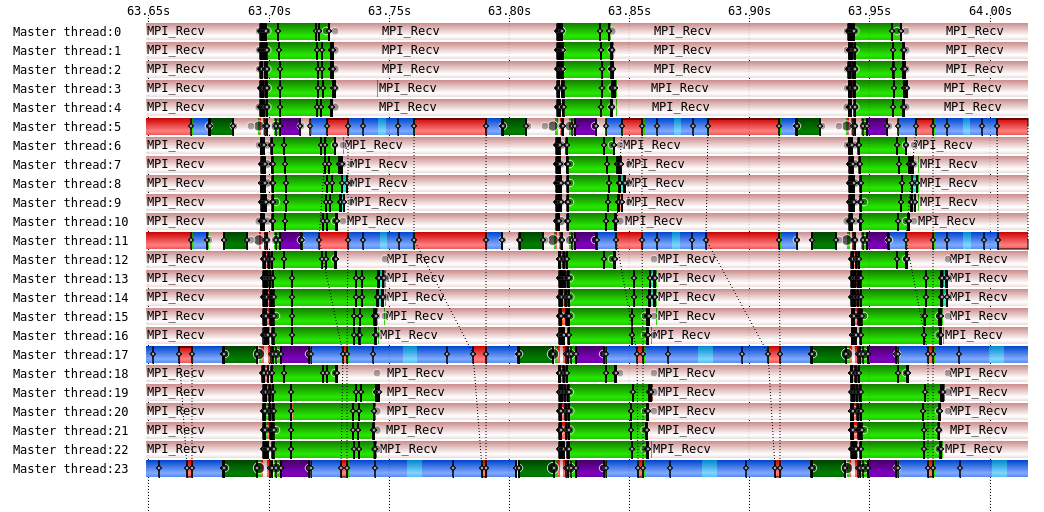}
    \caption{Tracing of a GROMACS run with 24 processes using ScoreP and Vampir tools. Four processes (three visible) complete the PME calculations in this run, while twenty processes quickly carry out the PP calculations. The PP processes wait for the PME processes to finish. We can also see a significant load imbalance between the PME ranks. The dark red is \texttt{MPI\_Alltoall} for FFT communication. All other MPI calls are light red. The green color represents the general compute, dark green is gather, blue is 1D FFT, purple is spread, and turquoise is the convolution calculation.}
    \label{fig:communication-gromacs}
\end{figure*}

As the first step of our study, we analyze the GROMACS parallel performance. Fig.~\ref{fig:communication-gromacs} shows a tracing of a GROMACS run instrumented with Score-P~\cite{10.1007/978-3-642-31476-6_7} and visualized with the Vampir tool\cite{knupfer2008vampir}. In this run, four processes (with thread numbers 5, 11, and 17) are dedicated to the PME calculations, while there are 20 PP processes. Within the PME processes, the all-to-all communication is reduced to only four processes decreasing the communication cost for the transposition in the 3D FFT.  It is important to note that the PP processes wait for the PME processes to finish in this run, and the PME calculations dominate the computational time step. We can also observe that the PME ranks are severely imbalanced as the 1D FFT calculations (blue) in rank 17 is much slower than the corresponding calculations in rank 5 and 11. 

\begin{figure}
    \centering
    \includegraphics[width=0.5\textwidth]{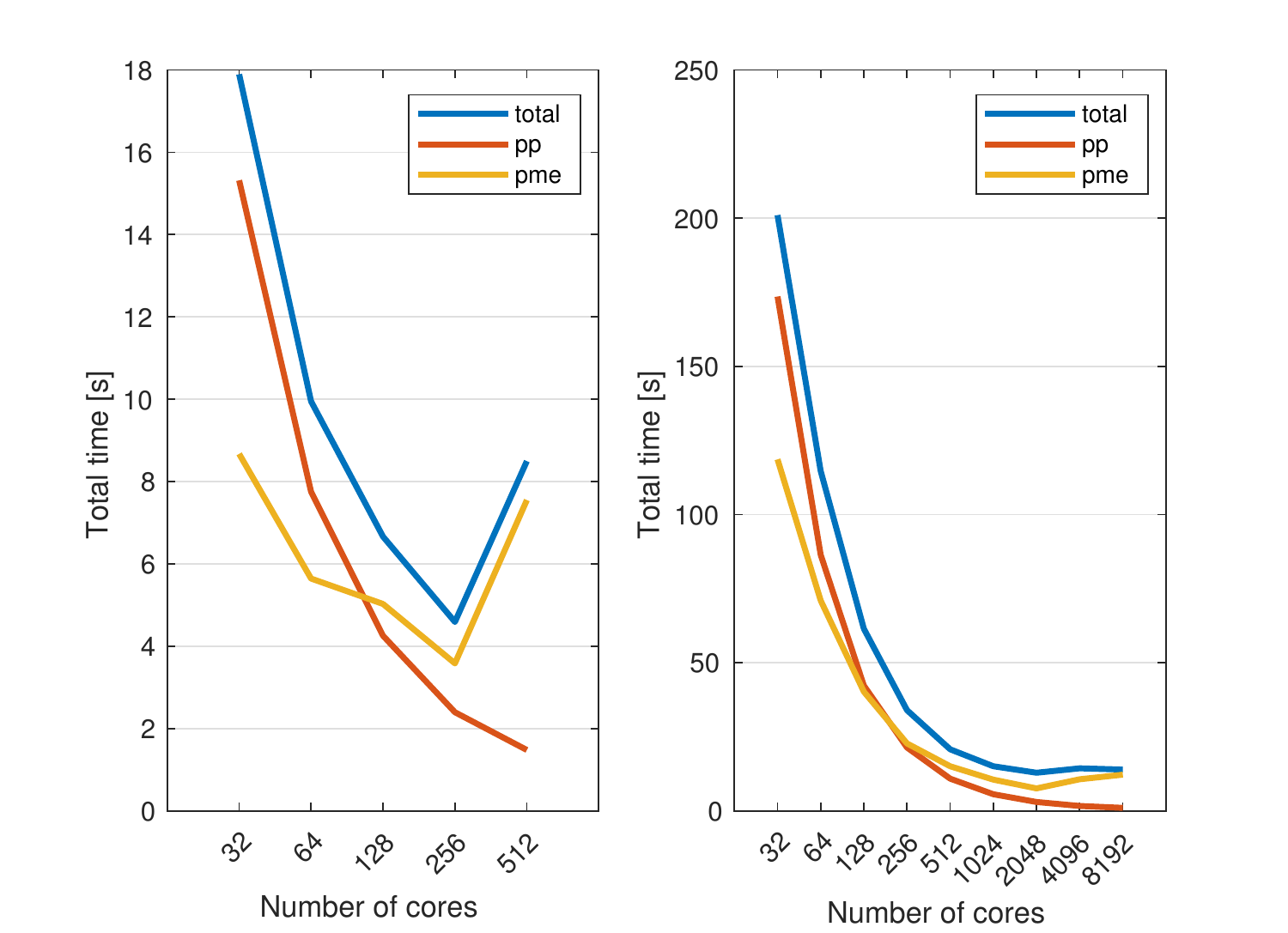}
    \caption{The fraction of time for PME, PP, and execution time on the Beskow system varying the total number of cores, to the left the Lysozyme system and to the right the Spike system. 1/4 of the cores are dedicated to the PME calculations. In this case, we use FFTW as the FFT library. Due to a lack of complete overlap and synchronization costs, the total execution is longer than each component phase duration.}
    \label{fig:pmefractionbeskow}
\end{figure}

We investigate the total impact of PME calculations on the GROMACS' total execution time. Fig.~\ref{fig:pmefractionbeskow} shows the fraction of simulation time spent in the PME calculations with respect to the PP time and the total time, varying the total number of cores per GROMACS simulation. 

The left panel presents the strong scaling results for the Lysozyme simulations (small-size problem). In this case, the simulation scales up to 256 cores then we observe an increase in the simulation time for 512 cores. In this case, more than 90\% of the simulation is spent on MPI communication. We note that the switch from a PP-dominated to PME-dominated simulations appears around 128 cores. In fact, PP scales well beyond 512 cores. There is a significant imbalance between PP and PME. This imbalance can be seen by inspecting the difference between the slowest PP and PME and the total time. The right panel of Fig.~\ref{fig:pmefractionbeskow} presents the results for the Spike strong scaling test. In this case, we observe strong scaling up to 2,048 cores. The simulation is bound by PP up to 256 cores and limited by PME beyond that. We note that PP keeps scaling like in the Lyso case. 

\begin{figure}
    \centering
    \includegraphics[width=0.5\textwidth]{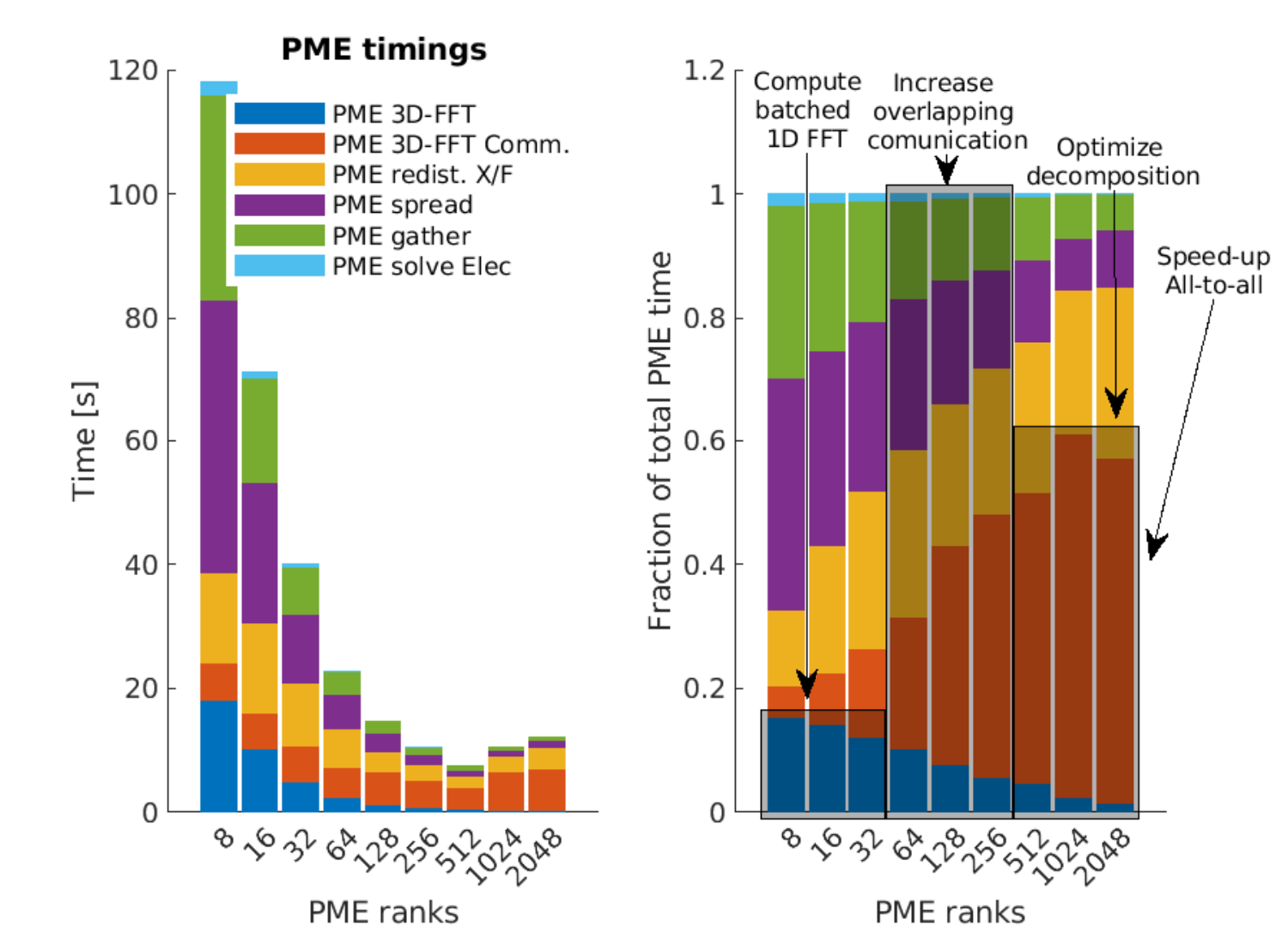}
    \caption{The left panel shows the strong scaling results for the different parts of PME for the Spike test case. The right panel present the PME time as a fraction of its components. The experiment was performed on Beskow, and the optimization phases are highlighted.}
    \label{fig:breakdownphase}
\end{figure}

After identifying PME as the main obstacle to strong scalability, we analyze which parts of the PME calculations show performance bottleneck and are amenable to performance optimization. Fig.~\ref{fig:breakdownphase} presents a breakdown of the different phases during the PME calculations on the Beskow system with FFTW (the right panel shows the percentage to ease the comparison). 

From an analysis of the plots, we note that when a small number of cores are dedicated to PME calculation, the PME spread and gather operations accounts for most of the time. On the contrary, for a larger number of cores, e.g., more than 256 PME cores, the communication for the 3D FFT (parallel transpose) and PME redistribution time dominate the PME calculation and, therefore, the whole simulation time. These two PME phases are responsible for losing scalability at large numbers of the core. Spread and Gather also level out at high core counts but at a lower total cost. At peak performance of the PME calculation, 50\% of the time is spent on 3D FFT calculation, and most of that is communication. 

\begin{figure*}
    \centering
    \includegraphics[width=0.49\textwidth]{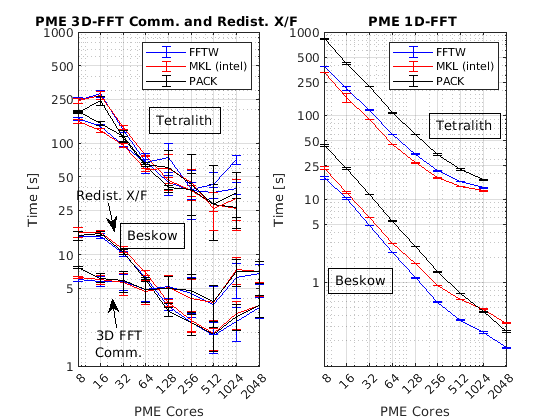}
    \includegraphics[width=0.49\textwidth]{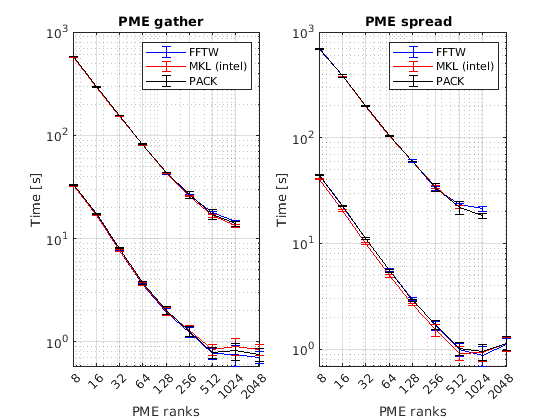}

    \caption{To the left: Strong scaling for the parts of PME with worst scaling on CPU for Spike on Beskow and Tetralith. To the right: The strong scaling of the FFT calculation. Note that the time (y-axis) does not align between plots.}
    \label{fig:scalingtestfft1}
\end{figure*}

An interesting question for GROMACS users is what performance improvement can be achieved by changing the FFT library and what is the best performing one in GROMACS. We compare the results for three PME phases (3D-FFT communication, PME 3D-FFT, and PME redistribution X/F) for different 1D FFT libraries. The results are shown in Fig.~\ref{fig:scalingtestfft1}.

As expected, we do not observe any significant performance change in the communication cost as GROMACS handles the communication, and it remains the same regardless of the library in use. Yet, we notice a difference in individual FFT performance. While for a small number of core counts, FFTW and MKL perform equally well, for a more significant number of cores, the FFT, built with GNU compilers, provides the best performance on Beskow. We also notice that the 3D FFT communication and redistribution measurements are noisy. We also note that the different clusters perform significantly differently: the Tetralith results communication results show a significant performance variability in Fig.~\ref{fig:scalingtestfft1}. We also note that MKL is slightly faster than FFTW. 

\begin{figure}   
   \centering
    \includegraphics[width=0.5\textwidth]{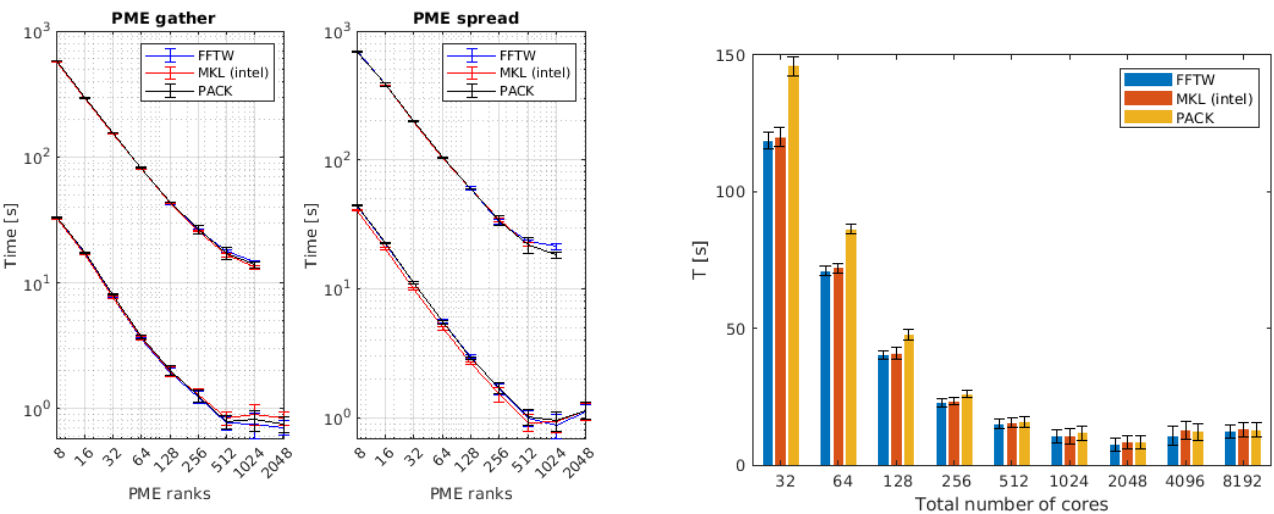}
    \caption{Strong scaling of the Gather (left) and Spread (right), top three lines are from Tetralith bottom three lines are from Beskow. The simulation uses 4th order interpolation.}
    \label{fig:Besvtet}
\end{figure}

We present the strong scaling behavior of the interpolation steps in Fig.~\ref{fig:Besvtet}. We notice that the scaling stops at approximately 512 cores for these parts, similar to the FFT communication and redistribution. However, it only displays a reduced performance variability after scaling breaks down compared with the parallel transpose and redistribution. We note that the gather and spread phases make up only 20\% of the total PME calculation after they have stopped scaling. 

\begin{figure}
    \centering
    \includegraphics[width=.49\textwidth]{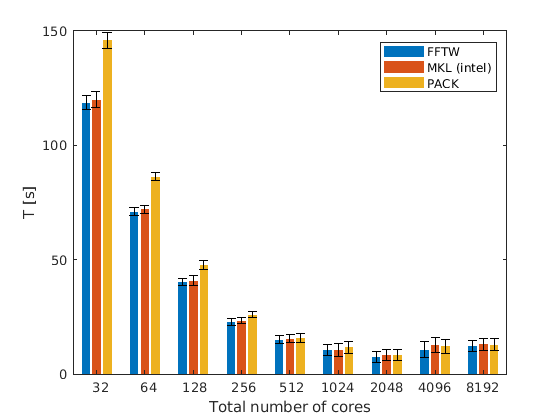}
    \includegraphics[width=0.49\textwidth]{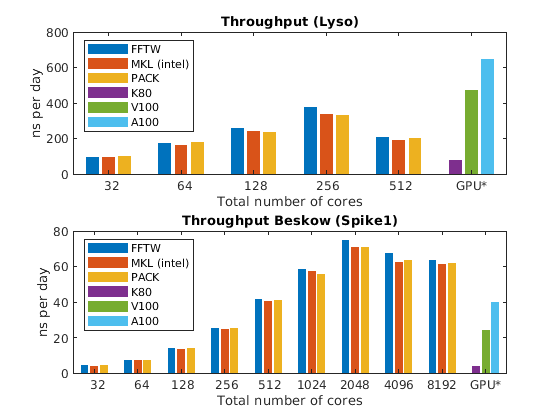}
    \caption{The total time for PME depending of the FFT library on Beskow with the Spike case.}
    \label{fig:pmetimefftlib}
\end{figure}

The total performance of the PME calculations, varying FFT libraries, can be seen in Fig.~\ref{fig:pmetimefftlib}. It shows the diminishing effects of the 1D FFT compute and the increasing performance variability coming from the communication with an increased number of cores. It is clear that FFTW and MKL are better choices than FFTPACK and are always motivated choices. 

\begin{figure}
   \centering
       \includegraphics[width=0.5\textwidth]{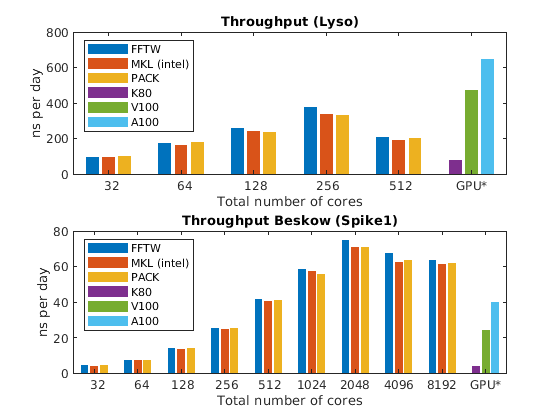}
   \caption{The total throughput for the different FFT libraries on Beskow (CPU), NJ (Nvidia A100 GPU) and Kebnekaise (Nvidia K80 and V100 GPU).}
    \label{fig:gpu}
\end{figure}

Finally, we analyze the performance of GROMACS PME on GPUs and present the results in Fig.~\ref{fig:gpu}. The GPU performance for the Lysozyme test case out-perform the CPU configurations significantly: PME on GPU and bonded calculations on CPU resulted in a throughput of approximately 640 ns/day compared with close to 380 ns/day for 128 CPU cores with FFTW.

We observe a slight advantage in using FFTW as the FFT library with a total throughput increase of 10\% compared with the other libraries for the small Lysozyme system. In the Spike test case, GROMACS runs on the Nvidia A100 perform comparably to 512 CPU cores on Beskow (46 ns/day) and a bit better than 512 cores on Tetralith (36 ns/day).

\section{Discussion and Conclusion}
\label{coclusion1}
In this paper, we presented an evaluation of GROMACS parallel performance. We conclude that the performance of PME is highly correlated with the parallel transpose and the redist functions. Redist is dependent on the domain decomposition of the PP part of the simulation, and therefore their performance is problem specific. The 3D FFT size can be varied to more accurately solve the Electrostatics or to balance the load. These two are also the parts of the PME calculation with the most variance between runs.  We identified three main factors in the GROMACS PME that can be improved. Firstly, there are many classes of problems, such as embarrassingly parallel ensemble jobs or parameter searching jobs, where many simulations can be distributed on many nodes with one job running on a single node. Improving 1D FFT performance for a single node can make noticeable overall performance in such ensemble simulations. Using FFTW or MKL instead of the backup library is advised; FFTW performs twice that of PACK in the spike problem. For the future, a possible further optimization technique is to use batched 1D FFTs parallelized with SIMD vector instructions.  Secondly, within the most scalable range, we can see that the cost associated with the interpolation steps and the transpose are similar in size. We would suggest an overlap between the communication needed from Spread-Gather and the parallel transpose. Since the GROMACS CPU code does not depend on an external 3D FFT library, it might be possible to incorporate the interpolation steps and therefore limit the need for communication. Finally, we have the transpose-dominated range -- the Achilles' heel of the method.

\section*{Acknowledgments}
{\small Financial support was provided by the SeRC Exascale Simulation Software Initiative (SESSI) and the DEEP-SEA project. The DEEP-SEA project has received funding from the European Union's Horizon 2020/EuroHPC research and innovation program under grant agreement No 955606. National VR contribution from Sweden matches the EuroHPC funding. The computations of this work were enabled by resources provided by the Swedish National Infrastructure for Computing (SNIC) at HPC2N, partially funded by the Swedish Research Council through grant agreement no. 2018-05973.

%
%
%
\bibliographystyle{IEEEtran}
\bibliography{main}
%

\end{document}